# Avalanche Photodetectors with Photon Trapping Structures for Biomedical Imaging Applications

CESAR BARTOLO-PEREZ[1*], SOROUSH CHANDIPARSI[1], AHMED S. MAYET[1], HILAL CANSIZOGLU[1], YANG GAO[1], WAYESH QARONY[1], AHASAN AHAMED[1], SHIH-YUAN WANG[2], SIMON R. CHERRY[3], M. SAIF ISLAM[1], AND GERARD ARIÑO-ESTRADA[3*]

[1]*Electrical and Computer Engineering, University of California – Davis, Davis, California, 95616, USA*
[2]*W&WSens Devices, Inc., 4546 El Camino, Suite 215, Los Altos, California, 94022, USA*
[3]*Department of Biomedical Engineering, University of California – Davis, Davis, California, 95616, USA*
*\*cbartolo@ucdavis.edu, garino@ucdavis.edu*

**Abstract:** Enhancing photon detection efficiency and time resolution in photodetectors in the entire visible range is critical to improve the image quality of time-of-flight (TOF)-based imaging systems and fluorescence lifetime imaging (FLIM). In this work, we evaluate the gain, detection efficiency, and timing performance of avalanche photodiodes (APD) with photon trapping nanostructures for photons with 450 and 850 nm wavelengths. At 850 nm wavelength, our photon trapping avalanche photodiodes showed 30 times higher gain, an increase from 16% to >60% enhanced absorption efficiency, and a 50% reduction in the full width at half maximum (FWHM) pulse response time close to the breakdown voltage. At 450 nm wavelength, the external quantum efficiency increased from 54% to 82%, while the gain was enhanced more than 20-fold. Therefore, silicon APDs with photon trapping structures exhibited a dramatic increase in absorption compared to control devices. Results suggest very thin devices with fast timing properties and high absorption between the near-ultraviolet and the near infrared region can be manufactured for high-speed applications in biomedical imaging. This study paves the way towards obtaining single photon detectors with photon trapping structures with gains above $10^6$ for the entire visible range.

## 1. Introduction

Many applications involving the detection of optical photons, such as quantum communications [1-3], light detection and ranging (LIDAR) [4, 5], and medical imaging [6-9] require high detection efficiency as well as a very fast timing response. In the case of time-of-flight positron emission tomography (TOF-PET), a nuclear medicine diagnostic technique, the time accuracy of its detectors is directly related to image quality [10]. State-of-the-art photodetectors currently used in TOF-PET scanners show single photon time resolutions (SPTR) of 100 ps, but devices with SPTRs as low as a few picoseconds are desired [11, 12] to allow to exploit the benefits of prompt-light mechanisms in gamma detector materials[13, 14].

Silicon (Si)-based photodetectors are very attractive due to their compact size, low production cost, and fine segmentation capability. However, Si-based devices offer a relatively low detection efficiency when fast photodetectors are designed with a thin absorption region. Likewise, a thicker detector designed for high efficiency contributes to poor timing performance. This effect, known as the bandwidth-efficiency product [15], represents an intrinsic challenge in semiconductor-based photodetectors.

Photon trapping micro and nanostructures in silicon have demonstrated the ability to modify the propagation of light from the incident direction to a perpendicular direction (Fig. 1). Such effects reportedly enhanced the absorption in both *PIN* (*p*-region, intrinsic-region, *n*-region) and *Metal-Semiconductor-Metal (MSM)* photodetectors at infrared wavelengths. Thus, the utilization of photon trapping nanostructures has been proposed to overcome the bandwidth-efficiency product limitation that semiconductor detectors pose, by enhancing the absorption and the probability of avalanche by electrons for higher gain and lower noise values [16-20].

In this work, we studied the improvement in the absorption of the optical power by photon trapping nanostructures in the visible wavelength range. Photon trapping nanostructures in silicon photodetectors have the potential to promote the initialization of avalanche by electrons, achieve higher multiplication gain values, and reduce the pulse time response, reducing the absorption in the highly doped regions, as illustrated in Fig. 1. Photon trapping structures were implemented in silicon photodetectors with 2.5 μm of thickness, to study their effect on absorption, gain, and time response. The external quantum efficiency (EQE) and gain were measured at 450 nm and 850 nm wavelengths. Finite difference time domain (FDTD) simulations were used to understand the absorption profile in the semiconductor, while electrical simulations complemented them by studying the electric field profile. We developed a simulation package that allowed us to optimize the design of photon trapping structures to achieve up to 90% of absorption at 450 nm wavelength for thin silicon layer of 1.2 μm. The combination of these results with an optimized doping profile is expected to contribute to the development of ultra-fast photodetectors with high gain and absorption efficiency. This was the first time that photon trapping structures demonstrate an enhancement in gain, absorption, and time response in silicon PDs at both ends of the visible range, opening the opportunity to develop highly sensitive receivers that could reach the single photon level, required for biomedical applications.

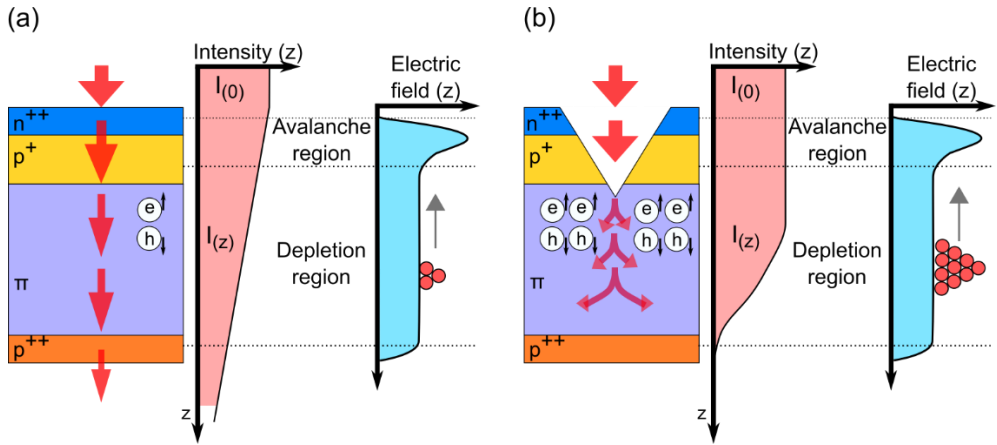

Fig. 1. Representation of the absorption and electric field profiles of two APD configurations. (a) Conventional APD (Control), and (b) Photon trapping APD (PT APD).

## 2. Device structure and fabrication

A silicon photodetector was fabricated in a mesa type structure with epitaxial layers grown with a total thickness of 2.5 μm on top of a silicon oxide insulator (SOI) substrate. An array of inverted pyramid holes was etched on the surface of the PD, serving as the photon trapping structures (PT PDs) as shown in Fig. 2(a) [right]. Two PDs with different period (*p*) and diameter (*d*) were studied (configuration 1: *p/d* = 630/900 nm and configuration 2: *p/d* = 1200/1500 nm). A PD with the exact same design and without any etched structures was also fabricated as a reference which is named the control PD [Fig. 2(a), left]. The doping profile of the silicon PD is described in the Fig. 2(b). The doping profile of the fabricated device created a *PIN* structure. Based on the design considerations, the *PIN* structure included a 2 µm intrinsic Si layer that was sandwiched between two ultra-thin (~0.25 µm) highly doped *n* and *p* layers. However, the *i*-layer thickness decreased to ~1.5 µm after the growth process due to carrier diffusion from high-doped layers. Further details of the fabrication methods are discussed in the literature [21].

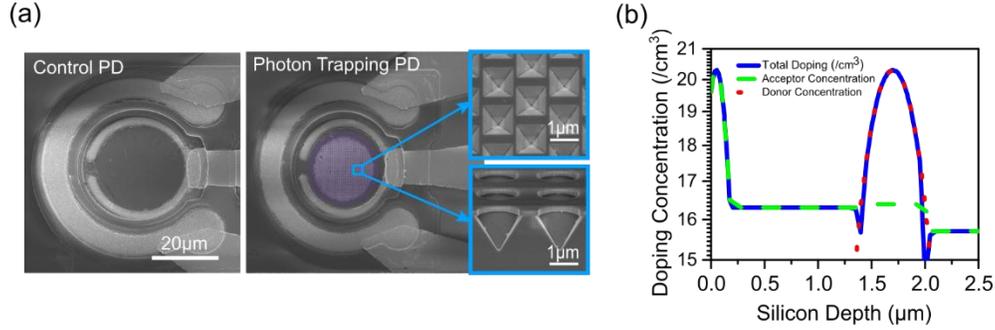

Fig. 2. (a) SEM image of control (left) and photon trapping (right) device. (b) Measured doping profile of the photodetectors.

## 3. Study using 850 nm wavelength

### 3.1 Optical-electrical simulation

The absorption profile of light with a wavelength of 850 nm was studied in a 2.5 µm-thick silicon photodetector using FDTD simulations by calculating the Poynting vector in each region of the semiconductor. The power absorption was then integrated with steps of 100 nm to understand where most of the absorption takes place. Lastly, the impact of this new absorption profile in the overall dynamics of the avalanche photodetector was studied by complementing the optical simulations with electrical simulations using an electrical solver software (Silvaco Inc, Santa Clara, CA, US).

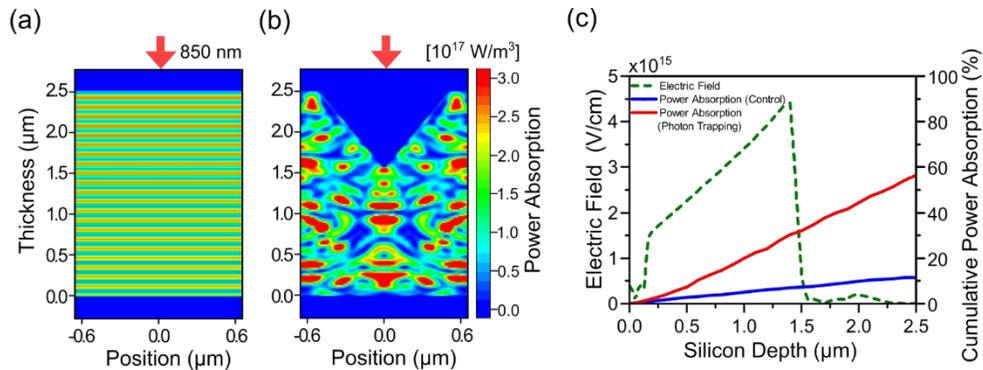

Fig. 3. Optical and electrical simulations in Si APD at 850 nm wavelength. Power absorption in (a) control Si PD and (b) PT-silicon PD. (c) Electric field profile of the fabricated device.

The simulated absorption profiles of control and PT PD are depicted in Figs. 3(a) and 3(b), respectively. At 850 nm wavelength, the control PD absorption was estimated to be 15%, as most of the input light was transmitted without interacting with the absorbing layer. The photon trapping device exhibited a distinctly higher absorption of 61% due to the enhanced light-matter interactions. The absorption profile suggests that it is feasible to control the injection of carriers in the avalanche photodetector to achieve higher signal-to-noise (SNR) ratios by increasing the gain and suppressing the excess noise. Figure 3(c) shows the calculated power absorption accumulated over the absorbing layer of the control and the photon trapping PD in steps of 100 nm. The electric field profile corresponding to the doping of the PD demonstrated that the absorbing region was completely depleted [superimposed in Fig. 3(c)]. Uniform high electric field over the depleted region ensures an improvement of the amplification factor by raising the impact ionization within the absorbing region.

### *3.2 Experimental gain and time response measurement at 850 nm wavelength*

Input light with a power of 8 µW at 850 nm wavelength was delivered to the surface of a photodetector with a diameter of 30 µm. The I-V curves shown in Fig. 4(a) describes a higher current in the photon trapping avalanche detector which was attributed to the enhanced absorption. The breakdown voltage in the control PD was measured as 34 V, whereas the photon trapping device showed a breakdown voltage at around 30 V.

The Multiplication Gain (M) was calculated as $M=[I_{photo(V)}-I_{dark(V)}]/[I_{photo(Vref)}-I_{dark(Vref)}]$, where $V_{ref}$ was taken at 10V. Multiple measurements were taking on the different devices on the same wafer, in order to consider the stochastic process of the avalanche process, obtaining a mean value <M> of 14.5 for the control PD and <M> of 554.6 for the photon trapping device [Fig 4(b)] with a standard deviation of ±0.6 and ±9.6, respectively. From the gain measurements we can identify the three regimes of operation of these PDs. Up to 10V the PDs present unity gain and hence considered PIN regime. Above 10V and below the breakdown voltage, where the gain values increase by the factor M, is considered the APD regime. Above their breakdown voltage, the devices operated in the Geiger mode regime and hence considered Single Photon Avalanche Detector (SPAD) mode.

The time response of the control and PT PDs was measured with a pico-second pulsed laser with 850 nm wavelength at the 3 regimens of operation: PIN (pink), APD (green) and SPAD (red) [Fig. 5(a) and (b)]. At 35 V, the SPAD regime, the photon trapping PDs exhibited a decrease in the FWHM from 99 ps to 40 ps, as well as a faster decay (fall time) from 293 ps to 105 ps of the pulse response. This can be attributed to the efficient delivery of the input light to the high electric field regions and a decrease of absorption in the highly doped regions, where diffusion is the dominant carrier transport method.

Most of the photons absorbed in the doped n and p regions of the photodiodes do not contribute to the photocurrents or EQE due to a lack of electric field in the highly doped p–n contact regions. A small fraction of photogenerated carriers in the photodiode contact regions are collected by the external circuit via the carrier diffusion process and contribute to the overall EQE. For an illumination wavelength of 850nm, ~1% of the light is absorbed in the top p region. However, when the top contact layer is thinned down from 200nm to 100nm, the percentage of absorbed light in the device's intrinsic region would increase by around 10% at shorter wavelengths such as 450nm. Farther thinning the top contact would augment the sheet resistance, contributing negatively to the bandwidth. We emphasize that the photon-trapping devices can inhibit surface reflection, absorb most photons, and exhibit reduced capacitance, contributing to higher absorption efficiency and high bandwidth. These devices also allow penetration of light to a much deeper level to maximize the gain, which is especially important for illumination with short wavelengths such as 450nm.

Junction capacitance and junction resistance, which form the device RC time constant, as well as the photogenerated carriers' transient time, are the main speed limiting factors in PDs. Photon trapping structures offer a solution to reduce the junction capacitance that depends on the device area and the depletion layer thickness. They could improve the absorption coefficient of the absorbing layer up to seven times[22] at 850nm, therefore, designing a small area device with similar absorption than large-area flat devices is feasible. Moreover, we have demonstrated that our photon-trapping structures can reduce the capacitance by more than 50% due to the reduction of the surface area of the device [23]. Transient time can be improved by employing thin highly doped contact layers (n++ and p++ layers) that facilitate the generated fast carriers' transition to the outer circuit elements and reduce the slow carries effect on device response.

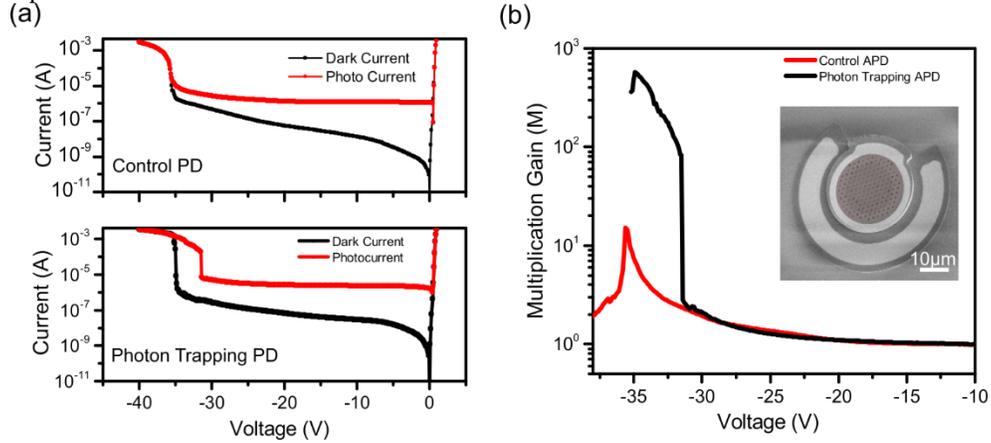

Fig. 4. Current-voltage gain and impulse response for Si APD. (a) I-V characteristics of control and photon trapping (PT) devices. (b) Multiplication gain of PT and control device.

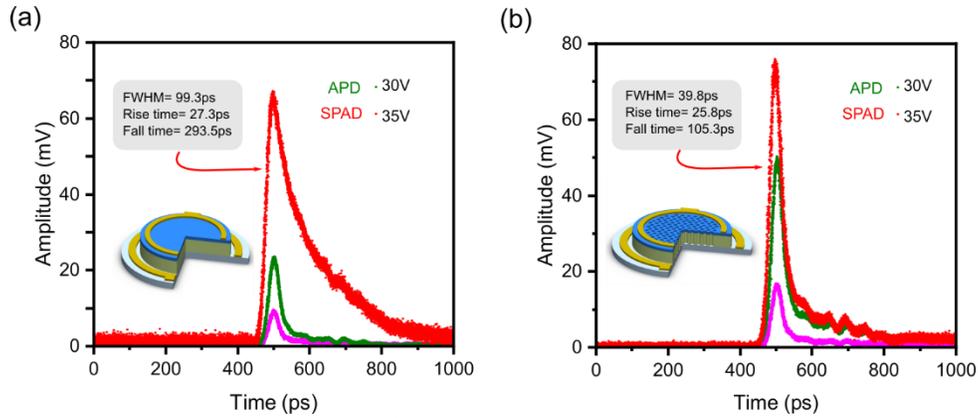

Fig. 5. Pulse time response for Si PD under the three regimes of operation PIN (pink) APD (green) and SPAD (red), for the (a) control and (b) Photon Trapping PD.

## 4. Study at 450 nm wavelength

### 4.1 EQE experimental evaluation

The EQE and gain of the fabricated photon trapping avalanche PD was measured at a wavelength of 450 nm. Two PT APDs devices with different diameter and depth of structures were characterized in addition to the control PD. The diameter of the two PT structures (holes) were 630 and 1300 nm with a calculated depth of 445 and 919 nm, respectively. Their depth was estimated according to the diameter of the structure and the characteristic angle of KOH etching of 54.74o. The EQE obtained for the photon trapping device with a diameter of 630 nm and a periodic distance of 900 nm was 82%, the PT PD with hole diameter of 1300 nm gave an EQE of 80%. On the other hand, the control device had an EQE of 54%. The EQE values obtained are corroborated by FDTD analysis. As described in Table 1, our PT device exhibit a responsivity of 297 mA/W at 450 nm and 425mA/W at 850 nm. Reported responsivities on silicon APDs range from 90 mA/W to 230 mA/W between 400 nm and 480nm [24-27] and from 4mA/W to 560 mA/W at 850nm[28, 29], as also reported in [30].

Table 1. Benchmark of different high speed silicon avalanche photodiodes

| Author/ Year | Material | EQE | Wavelength | Gain | Reference |
|---|---|---|---|---|---|
| Bartolo-Perez, 2021 | Si pin APD | EQE 82% (R=0.297A/W) EQE 62% (R=0.425A/W) | 450 nm 850nm | $M_{450nm}$=524 (27V) $M_{850nm}$=554 (29V) | This Work |
| Pancheri, 2007 | Si CMOS (0.35 µm) | EQE 23% (R=0.09 A/W) | 480 nm | 13 (10.3V) | [24] |
| Rochas, 2002 | Si APD 0.8µm | EQE 50% (R=0.16A/R) | 400 nm | 20 (19V) | [25] |
| Biber, 2000 | BiCMOS 2µm | EQE 25% (R=0.128A/W) | 635 nm | 7 (19.1V) | [26] |
| Pauchard, 2000 | custom | EQE 70% (R=0.237A/W) | 420 nm | 16 (14.1V) | [27] |
| Youn, 2014 | Si APD | 1.4 A/W with gain | 850 nm | (12.4V) | [28] |
| Stenindl, 2014 | High Voltage CMOS 0.35(um) | 0.41 A/W (75.8%) | 670nm | 6.6e$10^4$ $M_{optimum}$=50 | [29] |

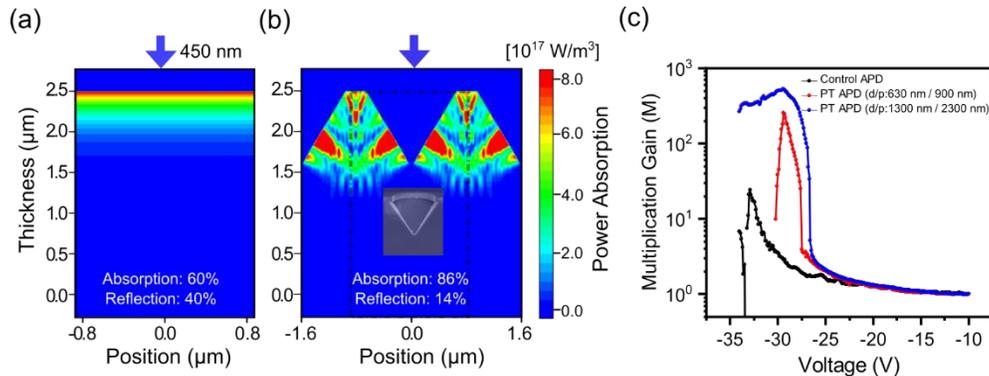

Fig. 6. FDTD simulations of 2.5 µm-thick APDs with input light of 450 nm for (a) a control APD and (b) a photon trapping APD. (c) Experimental gain measurements of fabricated devices at 450 nm wavelength.

As the FDTD simulations show in Fig. 6(a), in conventional silicon with a thickness of 2.5 µm, 60% of the light is absorbed, mostly close to the surface, while the remaining 40% is reflected. The implementation of the photon trapping structures reduces the reflection to 14% and the remaining 86% is absorbed [Fig. 6(b)]. In addition to the higher absorption capabilities of photon trapping silicon PDs, the multiplication gain was measured in the APDs with a laser diode emitting at 450 nm wavelength and a power of 30 µW. The bias voltage was swept from 1 to -34 V, in steps of 100 mV, for the devices with 30 µm diameter, to obtain the dark current and the photocurrent and calculate its gain. The control PD exhibited a gain of 24, while the PT PDs exhibited a gain factor of 157 and 524, for the 630 nm and 1300 nm diameter structures, respectively [Fig. 6(c)]. The difference in gain observed can be attributed to the different light penetration depths in the PDs. While in the control PD the 450 nm-wavelength light is absorbed close to the surface, the deeper PT structure increased the penetration length, affecting the probability of electrons and holes to generate an avalanche. The breakdown voltage also varies in the three devices. The PT PDs exhibit a breakdown voltage of 26-27 V, while the control PD required a voltage of 28 V. The variation in the breakdown voltage can be explained by the smaller net volume of the PT PD. The results of the photon trapping avalanche photodetectors studied at 450 nm wavelength are summarized in Table 2.

Table 2. Geometric details, and experimental measurements for EQE, gain, and breakdown voltage of control and photon trapping APDs for 450 nm wavelength photons.

|  | Control | Configuration 1 | Configuration 2 |
|---|---|---|---|
| **Shape** | - | Inverted pyramid | Inverted pyramid |
| **Diameter [nm]** | - | 630 | 1300 |
| **Depth [nm]** | - | 445 | 919 |
| **Periodicity [nm]** | - | 900 | 2300 |
| **EQE$_{450nm}$ [%]** | 54 | 82 | 80 |
| **Gain$_{450nm}$ [M]** | 24 | 157 | 524 |
| **Breakdown voltage$_{450nm}$ [V$_{BD}$]** | 28 | 26 | 27 |

*4.2 Simulation study for SPADs*

A series of optical and electrical simulations were performed to further optimize the design of photon trapping structures, with the aim to design Single Photon Avalanche Photodetectors (PT SPADs) that could enable ultrafast operation in the shorter wavelength of the visible range without compromising their sensitivity. In our proposed device, a silicon thickness of 1.2 μm was chosen to achieve low jitter time. A conventional device (control) with such a thin layer would be able to absorb only 51% of optical power at 450 nm wavelength, with most of the light being absorbed in the first 300 nm of depth, as FDTD simulations show in Fig. 7(a). However, our results suggest that more than 90% of photon absorption can be achieved in such a thin layer with optimized photon trapping structures, when a proper diameter and period is implemented in the photodetectors. Figure 7(b) shows the side and top view of the simulated PT SPAD suggesting a shift in the absorption with an enhancement peak at 400 nm depth just below the depth of the nanoholes.

Herein, we have studied the impact of design variations in period, diameter, and depth of cylindrical holes through FDTD simulation. Such simulations allowed us to construct Fig. 7(c), where the expected absorption for different diameters and periods, with a depth of 400 nm is presented. As observed in Fig. 7(c), the absorption efficiency shows a higher enhancement when the diameter of photon trapping structures approaches the period length between them, with a maximum of 90% absorption.

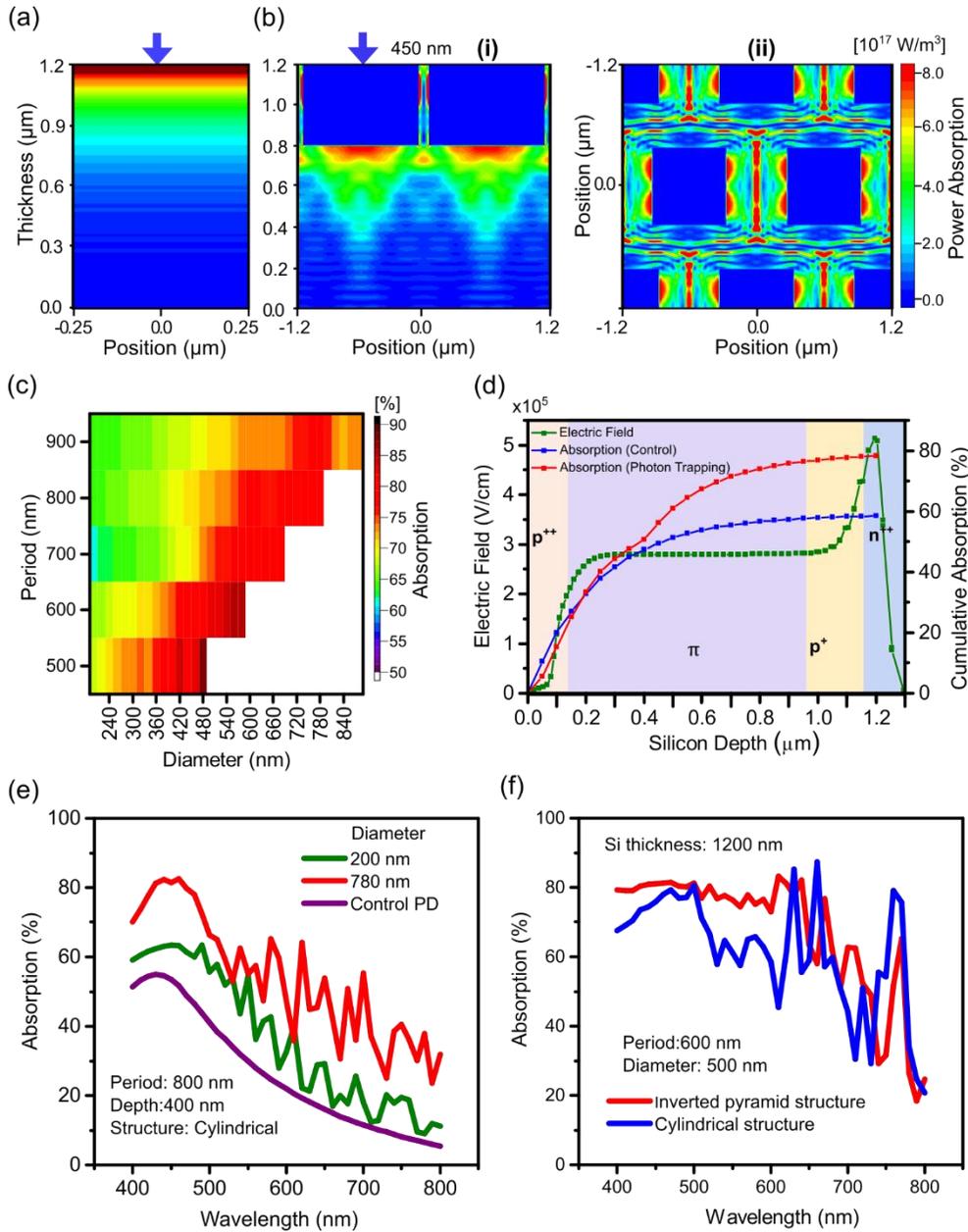

Fig. 7. Absorption control in photon trapping PD at 450 nm wavelength. Simulated power absorption profile of (a) control and (b) photon trapping PD with 1.2 μm thick silicon. Our photon trapping PDs with such a thin absorber layer exhibit more than 90% absorption. (c) Influence of period and diameter of the photon trapping nanostructures in power absorption at 450 nm wavelength. (d) Cumulative absorption in control (blue) and PT (red) silicon SPAD. Overlap of electric field profile of a PD with a *pπpn* structure with the absorption of light for optical generation, for higher gain and lower noise avalanche-based PD. (e) Influence of diameter in cylindrical photon trapping structure at broadband range of wavelengths. (f) Comparison of absorption at a broadband range of wavelengths between cylindrical and inverted pyramid structure.

Figure 7(d) shows the cumulative absorption obtained for the control (blue) and the PT SPAD device (red), in steps of 50 nm. In the first 100 nm depth, the high absorption observed in the control PD is reduced to almost 50%. Then, between the 400 nm and 500 nm depth, the absorption of power in a PT SPAD is increased by 3 times, compared with the control SPAD. This illustrates a reduction in optical power absorption close to the surface with the enhancement of absorption deeper in the device. With a proper electric field profile (green) that allows to separate the absorption and the multiplication regions, more electrons can be injected into the multiplication region, promoting higher gain and lower amplification noise [Fig. 7(d)].

The higher absorption efficiency achieved in the silicon-photon trapping APDs at visible and near-infrared wavelengths allows designing devices with thinner absorption layers. Such a reduction in thickness comes with a reduction of the breakdown voltage [31]. The breakdown voltage of our fabricated device with 2.5μm of thickness is around 30V and electrical simulations performed on the device proposed with 1.2μm of thickness, suggest a breakdown voltage of less than 20V. We envision a reduced voltage below 10V as a possibility in our devices with a thinner active layer.

An advantage of our photon trapping structures with respect to other absorption enhancement methods is its effect across a broad range of wavelengths, critical for biomedical imaging technologies [Fig. 7(e), (f)]. Figure 7(e) shows the higher absorption obtained in photodetectors with photon trapping structures from 400 to 800 nm. At 450 nm wavelength, the absorption increases as the hole diameter of the structure reaches close to the period within the structures. Different photon trapping structures can also be implemented, such as inverted pyramids or cylindrical holes. The simulated absorption of these structures reveals a higher absorption in the inverted pyramid profile from 400 to 600 nm wavelength [Fig. 7(f)]. In addition, in this range of wavelengths, a more constant absorption value is observed in the inverted pyramid structure. Engineering photon trapping structures in semiconductor-based photodetectors can benefit many biomedical applications that rely on the detection of optical photons with a broad distribution of wavelengths.

## 5. Conclusion

To the best of our knowledge, this study is the first comprehensive evaluation of gain, detection efficiency, and time performance of avalanche photodiodes (APD) with photon trapping nanostructures for photons of 450 nm wavelength. First, we have fabricated photon trapping photodetectors and evaluated their performance at 850 nm. Our detectors exhibited 30 times higher gain, 16% to >60% enhanced absorption efficiency, and a reduction in the FWHM of the pulse response of 50%, when operated in the SPAD regime. At an input wavelength of 450 nm, the EQE increased from 54% to 82% and the gain was 22 times greater in a device with photon trapping structures. Therefore, Si-based APDs with photon trapping structures showed a significant increase in absorption when compared to their flat counterparts. With an optimized doping profile and thinner layers, the gain can be further enhanced. Simulations at 450 nm show that 90% of the optical absorption can be obtained with only 1 μm of absorption layer thickness. This thin layer would significantly reduce the transit time of the generated carriers and improving time accuracy significantly. Such results show that thin devices with high absorption between the near-ultraviolet and the near-infrared region can be designed for ultra-fast operation.

**Funding**

This work has been funded by NIH grant R21 EB028398.

**Disclosures**

The authors declare no conflicts of interest.